# Siting and Curtailment Depth Limit the Grid Headroom Available to New Large Loads

Nils Johnson[1]*, Geoff Blanford[1], John E.T. Bistline[2,3]

**Affiliations:**

[1] Electric Power Research Institute; Palo Alto, CA, USA.

[2] Watershed; San Francisco, CA, USA.

[3] MIT; Cambridge, MA, USA.

*Corresponding author. Email: njohnson@epri.com.

## Abstract

Data center demand is increasing faster than generation can be built, encouraging proposals to interconnect large loads on condition that they reduce consumption during high-stress periods. Estimates of how much load can be added via flexibility rest on simplified accounting and report only annual curtailed electricity. Here we use a national energy systems model that co-optimizes hourly generation and load curtailment across 16 U.S. regions to characterize the flexibility and power system impacts of enabling headroom. While 50 GW can be served while curtailing less than 0.1% of its annual energy, small annual curtailment conceals potentially severe operating requirements. Most regions require curtailment deeper than 90% in at least one hour, so the load must disconnect or have sufficient on-site resources to provide full back up. Furthermore, announced data center locations are not aligned with modeled headroom and relying on existing generation fleets raises $CO_2$ emissions and wholesale electricity prices.

## Introduction

Electricity demand in the United States is increasing after two decades of little growth, driven primarily by data centers. Data centers could consume 380-800 TWh in 2030, with aggregate peak load rising by 24-71 GW between 2024 and 2030 [1]. Interconnection queues and multi-year lead times for new generation and transmission equipment leave data centers waiting years for grid connections [2], [3].

An increasingly attractive alternative to waiting is to connect large loads now, on condition that they reduce grid-supplied consumption during periods of high system stress. Because planning reserve margins are sized against a peak that occurs in a handful of hours, a load that cuts demand in those hours does not raise the peak and, in principle, does not require new capacity. The quantity of load that can be added in this way has been termed flexibility-enabled headroom [4]. Norris et al. estimated that 76-126 GW of new load could be integrated into U.S. systems if 0.25-1% of its annual energy were curtailed [4]. Curtailment can come from shifting or shedding compute, coordinating cooling systems, power-capping non-critical workloads, or substituting on-site energy storage and generation, several of which have been demonstrated at operating AI data centers [5]. Recent work has extended the headroom framing to generation adequacy, system cost, and resource adequacy planning [6]–[10].

Flexibility-enabled headroom has become a focus of industry efforts and policy. The AI Energy Management Alliance (AEMA) was launched in September 2026 to accelerate interconnection of flexible loads by a coalition of AI providers, data center operators, power producers, and grid operators, including Emerald AI, Google, and NVIDIA. Its stated priorities include defining obligations before a facility connects, standardizing performance metrics, and creating faster interconnection pathways for loads that make verifiable flexibility commitments [11], [12]. The coalition's claim is that flexible operation could accommodate roughly 100 GW on the existing U.S. grid [11]. Comparable proposals are under active consideration in regulatory proceedings on large-load interconnection [13].

However, despite planner and policy-maker interest, quantitative analysis of flexibility-enabled headroom remains limited, and estimates that anchor the policy debate rest largely on accounting using load-duration curves, which bounds what can be inferred in three ways. First, load-duration accounting reports the electricity that must be removed but not the operating regime that removes it. For instance, a curtailment obligation of 0.25% of annual energy could mean a single 22-hour outage or 175 hours at 12.5% depth. These scenarios imply different capabilities from the load's perspective, procured with different equipment. Second, existing studies estimate theoretical headroom under the assumption that new large loads will be sited where flexibility-enabled headroom is available, whereas actual siting decisions consider a range of factors, such as the costs and availability of land and water, proximity to fiber and labor, and speed to power. Finally, load-duration accounting does not track the implications of increasing the utilization of the existing fleet to serve load, which means that it cannot analyze the associated emissions and price impacts.

Here we address all three literature gaps using US-REGEN, a national energy systems model that co-optimizes hourly dispatch of generation, inter-regional transmission, energy storage and large load curtailment across 16 U.S. regions (Figure 1a). The analysis adds 10-150 GW of flat, high-load-factor demand nationally (on top of a 1,240 GW fleet) and solves two sets of scenarios at each level: FixCap, in which generation and energy storage capacity are frozen at 2024 levels so that all load must be met from existing assets plus curtailment; and OptCap, in which new capacity may be built. Curtailment is priced at $100,000 per MWh so that it is deployed only when the existing fleet cannot otherwise balance load during a given hour, which makes its dispatch a reflection of system demands for flexibility. In the FixCap scenarios, the model optimally allocates the siting of new large loads to regions that minimize required curtailment (hereinafter referred to as "flexibility-aligned siting"). In the OptCap scenarios, new large loads are allocated in alignment with the associated FixCap scenarios so that the power system impacts of restricting versus allowing new capacity can be compared.

While the use of a national energy systems model provides capabilities beyond those of load-duration accounting, it does not capture all system characteristics that could further constrain flexibility-enabled headroom and the resulting headroom estimates should thus be considered upper bounds. System-specific estimates would require the use of system-specific models and probabilistic resource adequacy assessment, which is beyond the scope of this analysis.

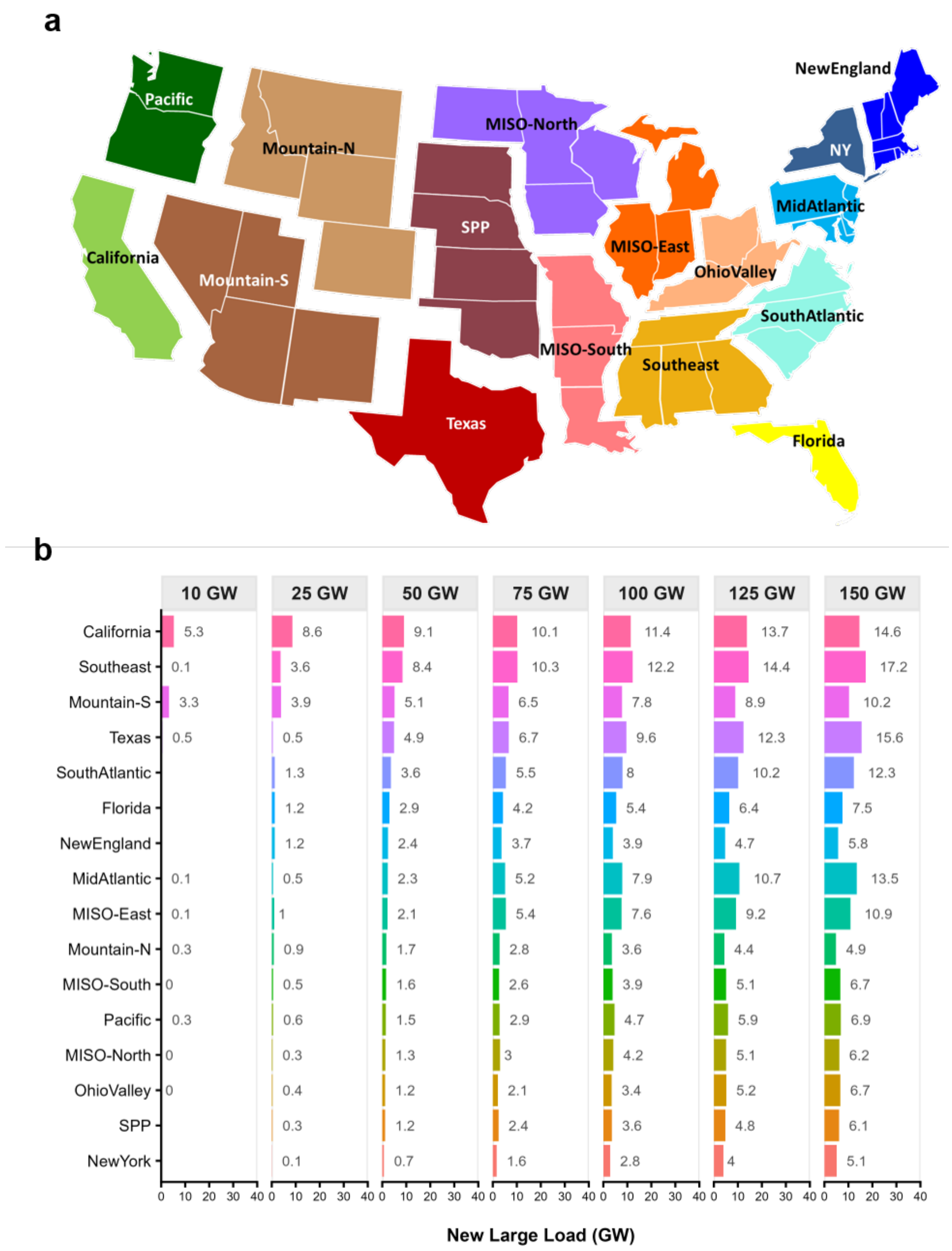

**Figure 1. Regional structure for REGEN analysis and headroom locations.** (a) The 16 model regions of REGEN used in this study. (b) Optimal regional allocation of new large load in the FixCap scenarios at each national load level from 10 to 150 GW. Regions are ordered by allocation at the 50 GW level. The first increments concentrate in California, Mountain-South and Texas, where existing energy storage contributed little to the 2024 reserve margin; at higher levels allocation shifts toward large systems with few near-peak hours, most notably the Southeast.

## Results

*Curtailment-free headroom is small and regionally concentrated*

One of the most influential factors dictating where new large load is initially assigned is the amount of existing energy storage capacity and the extent to which its discharge can be re-timed into the hour in which the reserve margin binds. Assuming that there is no excess generation capacity, the only options to add new large load without violating the minimum reserve requirement are to either deploy demand response (DR) or shift energy storage discharge into that binding hour. In our 2024 calibration, there are only three regions (California, Mountain-South, Texas) that have more than 1 GW of energy storage discharge capacity that can be re-timed in this way. When up to 10 GW of new large load is added, 90% of this load is allocated to these three regions (Figure 1b). No curtailment is required when up to 10 GW of new large load is added nationally.

At 25 GW, most new large load continues to be allocated to regions that can shift dispatch of energy storage to absorb its contribution to peak. However, as more load is added to these regions, storage becomes energy limited and its ability to integrate more load dissipates. At 50 GW and greater, the ability of energy storage to integrate new large loads while maintaining the minimum reserve margin becomes exhausted and the allocation of additional large loads is driven by three primary factors: 1. The size of the system; 2. The marginal cost of serving load from the existing regional fleet; and 3. The shape of the load profile. As new large loads increase from 50 to 150 GW, a growing share is allocated to regions with the largest peak demand, most notably Texas, Southeast, South Atlantic, and Mid-Atlantic. The growth in the allocation of large loads to the Southeast is particularly large because it has a "peaky" load profile, with relatively few hours in the top 10% of electricity demand. As a result, compared with other large systems, more new large load can be accommodated with less DR.

In this framework, flexibility-enabled headroom is therefore not a simple function of spare generation. Rather, it depends on how a region's energy storage, system size, and load shape interact, and some curtailment becomes necessary between 10 and 25 GW nationally, assuming that new large loads are sited in alignment with flexibility-enabled headroom.

*Annual curtailed electricity understates the load flexibility required*

To accommodate up to 50 GW of new large loads nationally, assuming siting aligned with flexibility-enabled headroom, total curtailed electricity remains less than 0.1% of the annual demand from these new loads (Figure 2). Required curtailment reaches about 1.5% of the annual electricity demand of these new loads when 150 GW are added nationally (ranging from 0.9% in California to 2.0% in the Ohio Valley). On this metric, our results are very similar to those reported by Norris et al. [4], as shown in Figure 2. However, a seemingly small share of annual load curtailed masks the operational implications of curtailment, which is characterized by the depth, frequency, and duration of individual curtailment events.

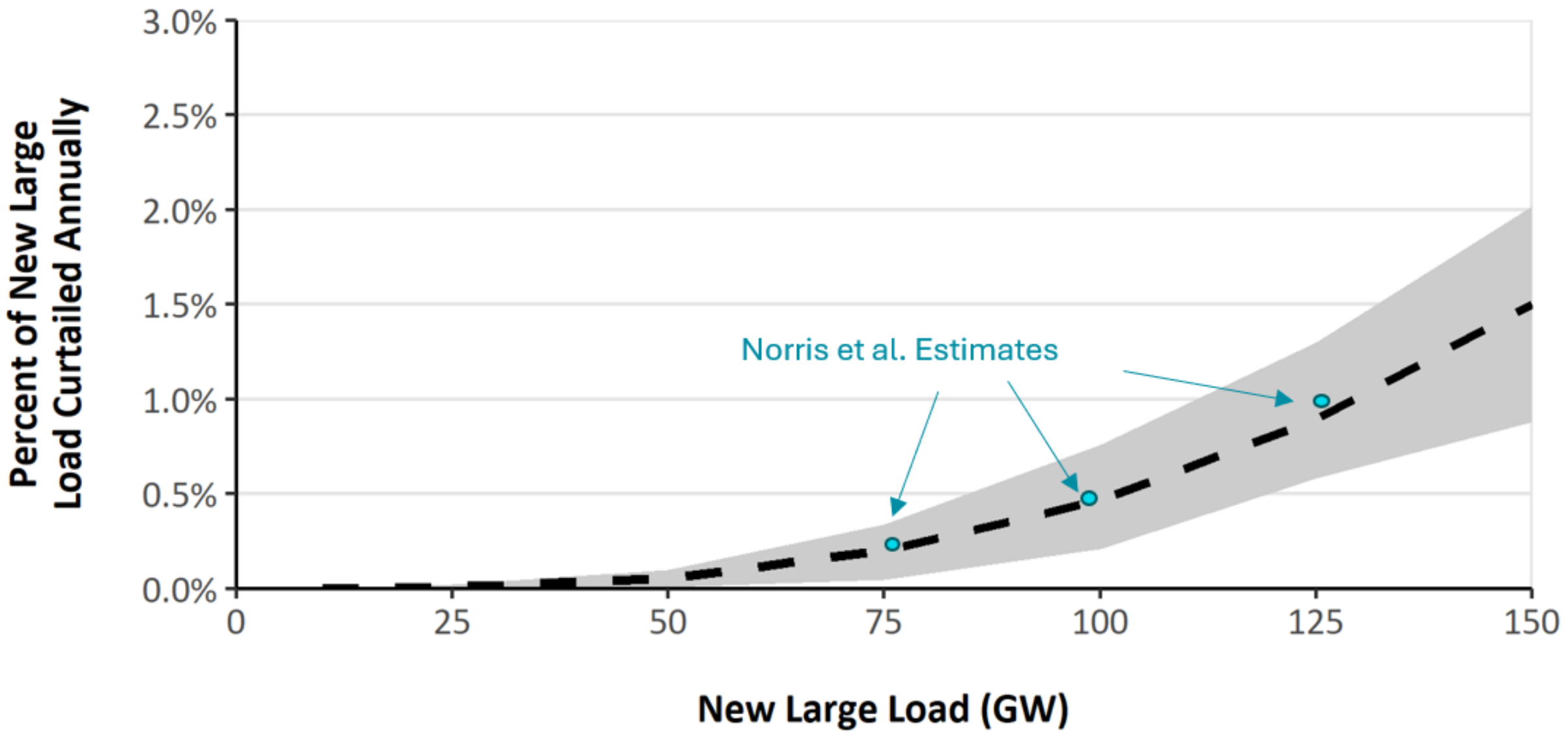

**Figure 2. Annual curtailment required to integrate new large load with existing capacity.** Percentage of the annual electricity of new large load that must be curtailed, as a function of national new large load, under FixCap with siting aligned with flexibility-enabled headroom. The dashed line is the load-weighted national average, and the grey band spans the range across the 16 REGEN regions. Points show the estimates of Norris et al. [4] for comparison.

Table 1 and Figure 3 provide a more complete characterization of the DR required to supply each amount of new large load from existing capacity. In flexibility-aligned regions, up to 10 GW can be added with no DR. Integrating 25 and 50 GW requires DR in no more than 20 hours of the year in any region, across at most 14 events, and no event exceeds eight hours (Table 1). However, 7 and 14 of the 16 regions require more than 90% of load to be curtailed in at least one hour when 25 and 50 GW are added, respectively, and several regions require 100% curtailment in at least one hour. Thus, even when only 25 GW is added, the large loads in many regions would have to curtail or have on-site generation to cover nearly all their peak load in at least one hour.

These depth requirements imply stark choices for facilities. A requirement to curtail essentially all load in an hour can be met either by disconnecting or by covering the load with on-site generation or energy storage. Disconnection forfeits the output of the facility for the duration of the event, and the 25% curtailment for three hours demonstrated at one operating AI data center is considerably shallower than the depth of curtailment required in this analysis [5]. On-site capacity avoids that loss, but because it would only run for a few hours annually, its levelized cost of electricity is very high, and the capacity may not necessarily be avoided but relocated behind the meter, drawing on the same constrained equipment supply chains motivating flexibility in the first place.

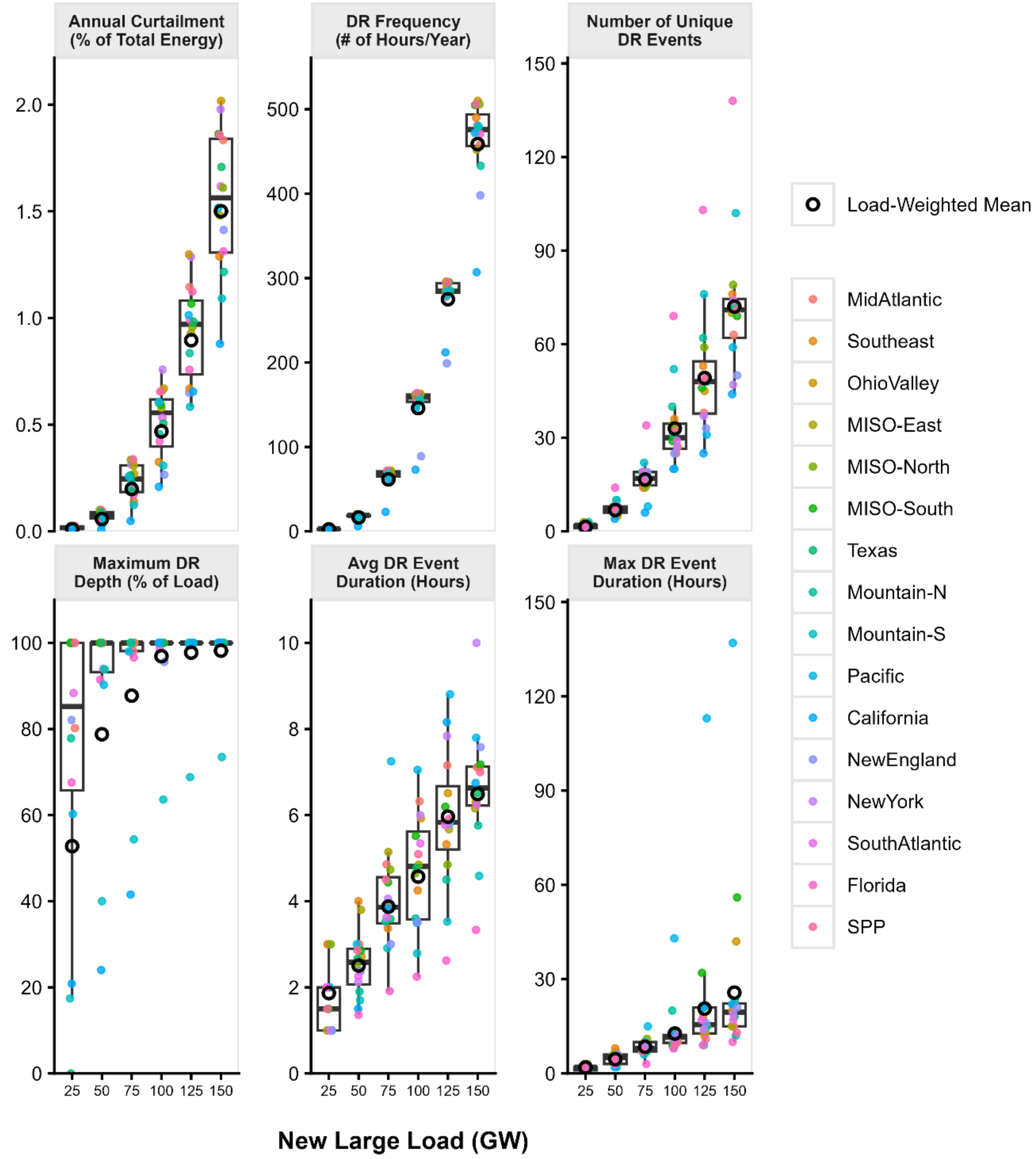

**Figure 3. Operational characteristics of the required curtailment.** Regional distribution of six curtailment metrics in the FixCap scenarios at each national new large load level: annual curtailment (% of new-load energy); curtailment frequency (hours per year); number of independent curtailment events; maximum curtailment depth (% of regional new large load); and mean and maximum event duration (hours). Colored points are individual regions; open black circles are load-weighted national means; boxes show median and interquartile range across regions. The 10 GW case is omitted because no curtailment is required.

**Table 1. Characteristics of the large-load curtailment required to integrate each national increment of new large load under the FixCap scenario.** Average denotes the load-weighted national mean; max and min are the extreme values across the 16 regions. Frequency is reported as the total number of curtailment hours and as the number of hours exceeding depth thresholds of 25%, 50%, 75% and 90% of regional new large load. The 10 GW case is omitted because no curtailment is required.

| Metric | 25 GW | 50 GW | 75 GW | 100 GW | 125 GW | 150 GW |
|---|---|---|---|---|---|---|
| Curtailed energy (% of new-load annual energy) | <0.1 (<0.1–0) | <0.1 (0.1–<0.1) | 0.2 (0.3–<0.1) | 0.5 (0.8–0.2) | 0.9 (1.3–0.6) | 1.5 (2.0–0.9) |
| Maximum curtailment depth (% of new load) | 53 (100–0) | 79 (100–24) | 88 (100–42) | 97 (100–64) | 98 (100–69) | 98 (100–74) |
| Regions with >90% maximum depth (% of regions) | 44 | 88 | 88 | 94 | 94 | 94 |
| Curtailment hours, depth >90% | <1 (2–0) | 2 (6–0) | 2 (7–0) | 3 (8–0) | 5 (10–0) | 8 (17–0) |
| Curtailment hours, depth >75% | <1 (2–0) | 2 (6–0) | 4 (10–0) | 9 (18–0) | 15 (35–0) | 22 (47–0) |
| Curtailment hours, depth >50% | <1 (2–0) | 4 (9–0) | 11 (27–0) | 26 (55–4) | 50 (104–12) | 82 (155–30) |
| Curtailment hours, depth >25% | 1 (3–1) | 7 (16–0) | 27 (48–8) | 66 (108–25) | 125 (178–81) | 217 (284–118) |
| Curtailment hours, total | 2 (3–2) | 16 (20–6) | 62 (72–23) | 146 (164–73) | 275 (296–199) | 459 (510–307) |
| Independent curtailment events | 1 (3–1) | 7 (14–4) | 17 (34–6) | 33 (69–20) | 49 (103–25) | 72 (138–44) |
| Mean event duration (h) | 2 (3–1) | 3 (4–1) | 4 (7–2) | 5 (7–2) | 6 (9–3) | 6 (10–3) |
| Maximum event duration (h) | 2 (3–1) | 5 (8–2) | 8 (15–3) | 13 (43–8) | 21 (113–9) | 26 (137–10) |

Unlike depth, required frequency and duration of DR keep growing at higher levels of new large load. At 75 GW, DR is required for an average of 62 hours per year across 17 independent events with a mean duration of 4 hours. At 100 GW, this rises to 146 hours over 33 individual events. To integrate 125 or 150 GW of new large loads, DR is required in hundreds of hours with the longest events in some regions lasting several days (Table 1). While the percentage of annual energy curtailed remains below 1% when even 125 GW of new large load is added to the current system, the depth, duration, and frequency of the required DR illustrate the operational challenges of integrating new large loads with only existing capacity and flexibility.

*Announced siting is misaligned with modeled headroom*

Actual siting decisions for new data centers and other large loads, while strongly influenced by the availability of power, are also based on other factors such as access to land and water resources, proximity to labor markets, and other infrastructure such as the fiber optic network. Based on announced locations of projects, EPRI's medium projection estimates about 71 GW of realized peak load from data centers in 2030, of which 49 GW are new between 2024 and 2030 [1]. Of that 49 GW, roughly 40% is concentrated in two regions, Texas and South Atlantic, which includes Virginia (Figure 4a).

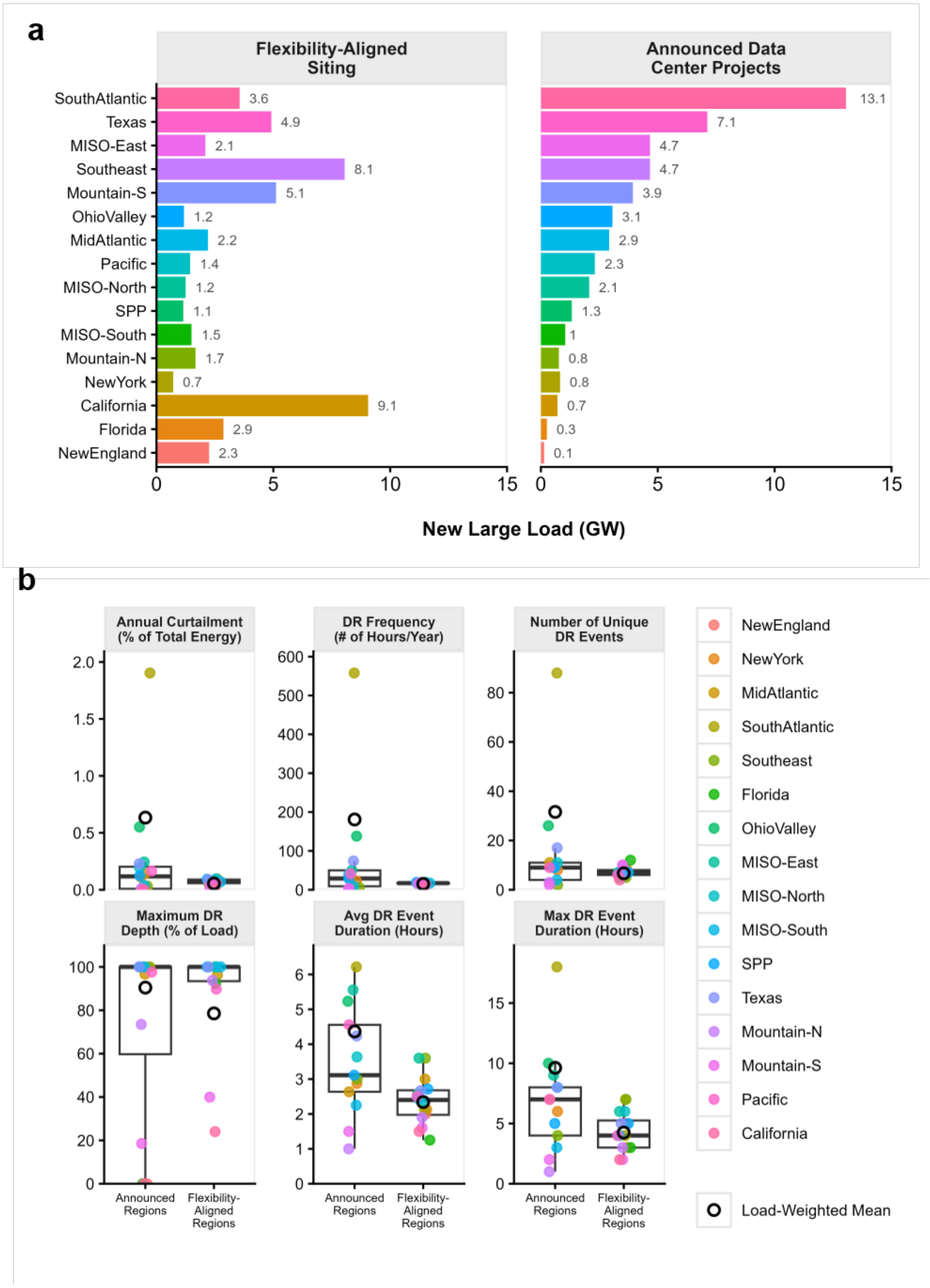

**Figure 4. Announced data center siting versus flexibility-aligned siting.** (a) Regional allocation of 49 GW of new data center peak load projected for 2030 under EPRI's medium scenario, distributed according to project announcements (left) and according to flexibility-enabled headroom (right). (b) Regional distribution of the six curtailment metrics of Figure 3 for the two regional allocations. Colored points are individual regions, open black circles are load-weighted national means.

We compare serving that data center load in its announced regions with allocating the same 49 GW according to modeled headroom. In both cases, more than half of new data center loads are concentrated in only four regions, but the regional allocation is quite different. In particular, the South Atlantic region is projected to have about 13 GW of new data center aggregate peak load based on announcements but is allocated less than 4 GW when considering flexibility-enabled headroom. Of the top four regions in the Announced scenario, only Texas and Southeast are also in the top four in the Flexibility-Aligned scenario.

For all DR metrics, the variance across regions is much smaller in the Flexibility-Aligned scenario (Figure 4b). The national load-weighted averages for all metrics are significantly larger in the Announced scenario, but the larger values are driven primarily by very high DR requirements in the South Atlantic region. In the absence of new capacity, about 2% of the annual energy associated with projected data center loads in 2030 would be curtailed in South Atlantic (i.e., more than 550 hours across 88 events), while 0.6% would be curtailed in Ohio Valley. In contrast, when new data center loads are allocated to regions according to flexibility-enabled headroom, curtailment of annual energy is less than 0.1% in all regions.

Under a curtailment ceiling of 0.1% of annual energy, six regions cannot meet their projected 2030 data center load when relying on existing capacity and flexibility alone: South Atlantic, Texas, MISO-East, Ohio Valley, Pacific, and MISO-North. In contrast, no region exceeds that ceiling when the same national total is sited according to headroom. Thus, announced new data center loads are not always aligned with modeled headroom which suggests that the practical potential of flexibility-enabled headroom to integrate new large loads is diminished relative to the Flexibility-Aligned scenario. Moreover, even the announced data center load that could be integrated with limited curtailment would still require curtailment deeper than 90% in at least one hour in most regions.

*Constraining new capacity shifts generation toward the incumbent coal and gas fleet*

When new large loads are introduced in the OptCap scenarios, a diverse mix of new capacity is deployed including solar, wind, and natural gas technologies to complement existing resources (Figure 5a). When 25 GW of new large load is introduced, only 11.3 GW of new solar, 0.3 GW of wind, and 7.5 GW of natural gas capacity are built nationally. Beyond 50 GW, investments are increasingly dominated by natural gas turbines and combined-cycle capacity which provide both reserve capacity and relatively low-cost firm generation. In the OptCap scenarios, the higher efficiency of new natural gas combined-cycle capacity moves this capacity higher in the merit order, displacing generation from existing coal- and gas-fired capacity seen in the FixCap scenarios (Figure 5b).

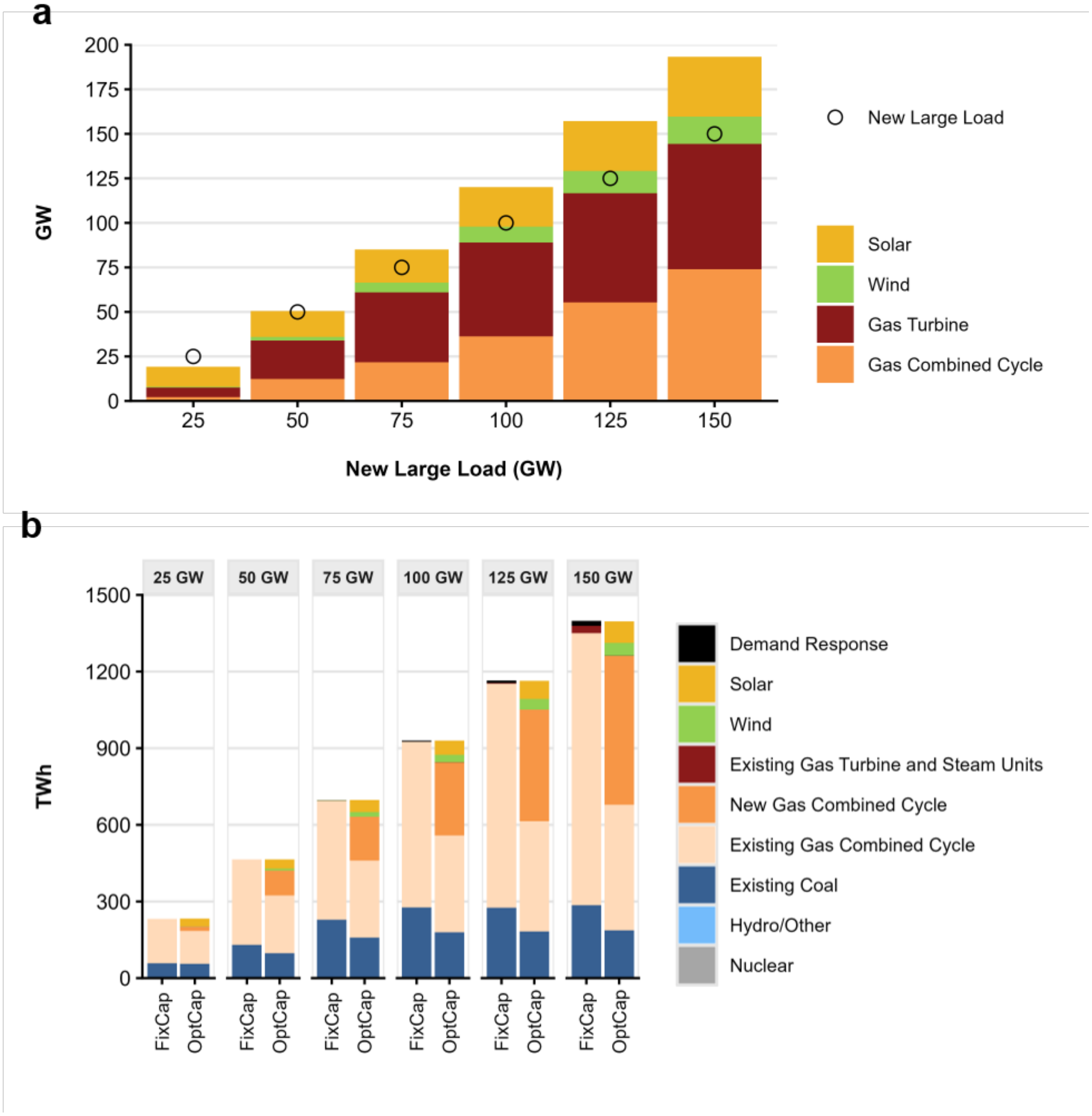

**Figure 5. National capacity and generation responses to new large load.** (a) Incremental capacity additions by technology in the OptCap scenarios (circles show the new load served). (b) Incremental generation by technology and resource in the FixCap and OptCap scenarios across large load levels.

Implications for emissions and prices follow from these dispatch changes. The average $CO_2$ intensity of incremental generation serving the new load exceeds that of the existing grid in both cases but is up to 24% lower under OptCap than FixCap scenarios (Figure 6a). The higher emissions intensity for serving new large loads in the FixCap scenarios is caused primarily by the greater reliance on existing coal-fired generation and, to a lesser extent, by reliance on less efficient existing natural gas combined-cycle capacity. Note that these changes exclude potential emissions associated with DR, which could be provided via on-site backup generation using carbon-intensive small-scale gas turbines and/or diesel or gas-fired spark-ignited engines, if permitted to do so. Increased utilization of coal- and gas-fired generators would also negatively impact air quality in neighboring communities.

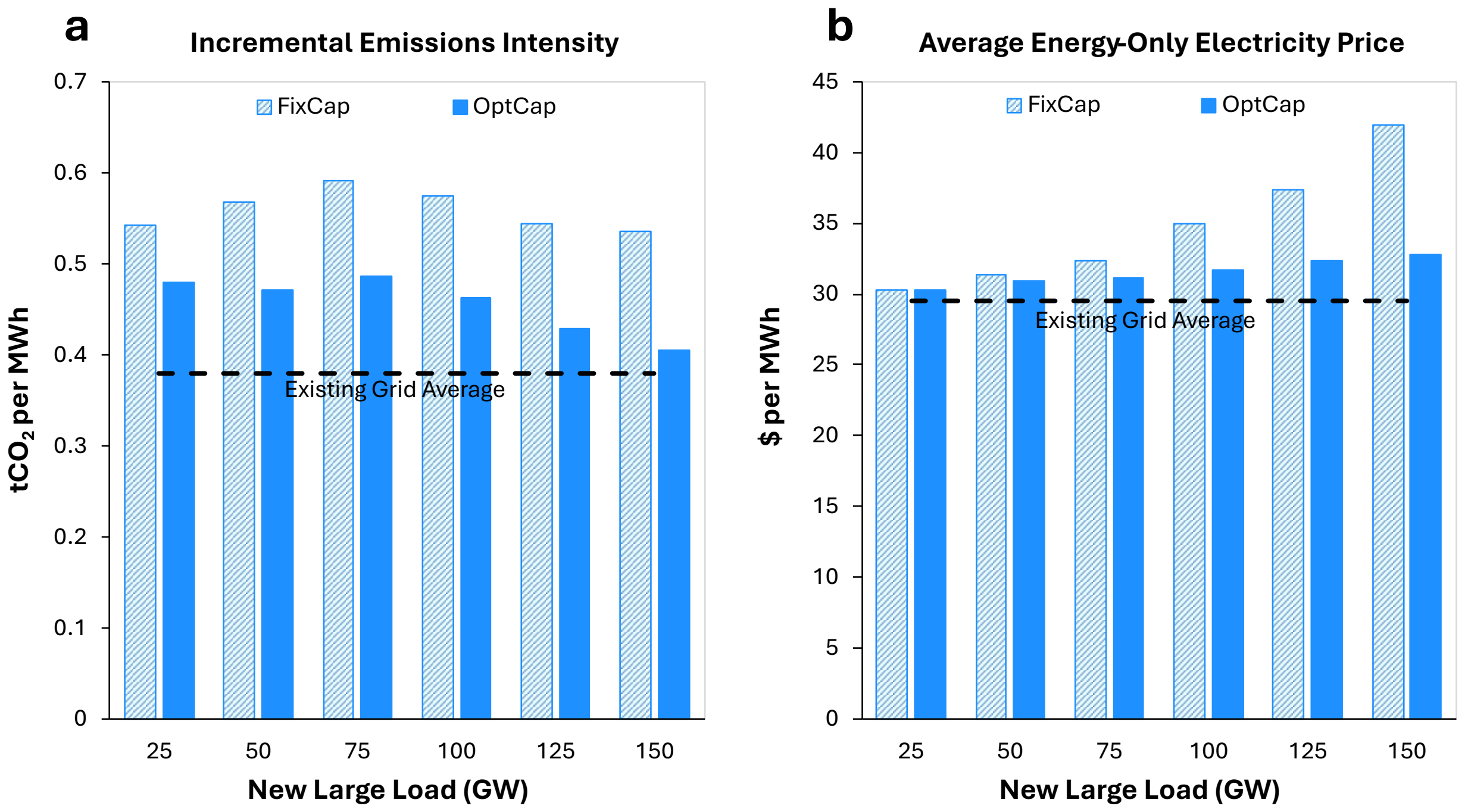

**Figure 6. Emissions and price implications of constraining new capacity.** (a) $CO_2$ emissions intensity for incremental supply to meet new large load. Dark blue bars are the OptCap scenarios, and light blue bars are the FixCap scenarios. Emissions from behind-the-meter generation used to deliver curtailment are not included. (b) National average energy-only wholesale electricity price (2023 USD/MWh) under FixCap and OptCap scenarios.

While the energy-only component of the wholesale electricity price (reflecting short-run marginal cost) remains relatively stable in response to higher demand in the OptCap scenarios, the price in the FixCap scenarios is much more sensitive (Figure 6b). At 50 GW of new large load, the energy-only price begins to increase relative to the corresponding OptCap scenario and the difference becomes larger as more new load is introduced (4% increase at 75 GW and 28% higher at 150 GW). Meeting new loads while relying on existing resources and flexibility alone can lead to both higher wholesale electricity prices and higher emissions since greater dispatch of generators with higher $CO_2$ intensities and dispatch costs are required as more new large load is added.

## Discussion

This analysis quantifies the implications of flexibility-enabled headroom in the U.S. under alternate levels of new large loads. It shows that useable headroom is smaller, more dependent on location, and more operationally demanding than implied by energy-based estimates from earlier studies. Three findings provide insights relevant to electric sector planning, tariff design, and policy-making.

First, annual curtailed energy is an incomplete metric for assessing feasibility. It is the statistic that accounting based on load duration curves naturally produces, but it says little about the characteristics of the flexibility that facilities would have to provide. Notably, curtailment depth saturates near 100% when relatively small levels of new large load are added even though annual curtailed energy remains small, which implies that the transition from “no flexibility required” to “must be able to drop to zero” occurs very quickly. Flexible load tariffs and interconnection agreements specified in terms of annual energy will therefore understate what would be asked of the load. More informative metrics include specifications of depth, frequency, duration, and maximum event length (quantities highlighted in Table 1).

Second, headroom is a location-specific property, and national totals are sensitive to the adequacy criterion applied, which headline figures rarely state. The roughly 100 GW figure currently being discussed in advocacy for flexible interconnection [11] is about twice the national total our idealized cases reach considering practical limits on curtailment and power system impacts, and the gap widens once load is placed where it is being proposed. This is a difference between how much load a national fleet could theoretically support and how much it could support where load is being planned (i.e., without spatial flexibility in allocating the new load). Policies that reward flexibility without regard to location will deliver a fraction of the headroom that motivated them.

Finally, avoided capacity does not necessarily mean avoided cost. Under scenarios fixed to the current fleet, the load must be able to cover or forgo its own peak, which means that the facility either forgoes compute-related income or deploys capacity behind the meter, where it is less utilized and, per unit of firm capability, more expensive. Meanwhile, the system absorbs higher emissions and higher energy prices as it is forced to rely increasingly on generators that are lower in the merit order. A complete comparison of demand- and supply-side strategies requires costing the behind-the-meter capability and its emissions alongside the utility investment it displaces, which our framework does not do and which we regard as the priority for subsequent work.

The analysis suggests that emissions and short-run marginal electricity prices can rise if capacity is fixed in meeting new large loads, though the extent depends on the scale of the new load. These system effects require systems modeling to capture, which was not addressed in earlier studies based on load duration curves. Other studies of tradeoffs between capacity expansion and fixed fleets for meeting data center load also suggest that constraining new capacity can have adverse impacts for emissions and cost, though the extent depends on system conditions [8].

This paper identifies considerations that assessments of flexibility-enabled headroom should address, but a system-specific estimate of useable potential requires more detailed analysis than is provided here. Omitted dynamics include intra-regional transmission and distribution (only inter-regional transmission is captured), plant-level operating constraints, multiple weather and load years, imperfect foresight (especially during system stress periods), and other flexible demands (Methods). Each omission suggests that the headroom we report should be considered an upper bound.

## Methods

### *Model*

This study is conducted using EPRI's US-REGEN model, an economy-wide energy systems model of the continental U.S. with a detailed representation of the power system; end-use sectors; and fuel transformation, storage, and delivery. Detailed model documentation, recent peer-reviewed journal articles, and a link to the open-source repository can be found at us-regen-docs.epri.com.

The version of the model used in this study has 16 regions (Figure 1a) and jointly optimizes capacity investment (for applicable scenarios) and hourly dispatch of generation, inter-regional transmission, and energy storage technologies for a single year, subject to hourly energy balance and planning reserve margin constraints. In addition to supply-side dispatch, the model optimizes hourly dispatch of any large load curtailment required to ensure system balancing with the integration of new load.

To assess the magnitude of potential headroom (i.e., new load that can be reliably served with existing capacity), it is important to align the modeled existing generation and energy storage capacity with available data. The total installed power generation capacity in the continental U.S. in 2024 is calibrated using Energy Information Administration (EIA) Form 860 data accessed through the Hitachi Energy Velocity Suite as of January 2025 [21]. According to this dataset, there was about 1,240 GW of net summer generation and energy storage capacity in the continental U.S. at the end of 2024. The study also uses hourly load profiles for 2024 from FERC Form 714 accessed through the same source. Hourly renewable profiles are also based on the 2024 weather year. This study does not consider the sensitivity of headroom and flexibility needs across multiple weather years or potential future load profiles.

US-REGEN includes a minimum reserve margin constraint, which requires that extra capacity is available above the forecasted peak load in a power system. The reserve margin is essentially the safety net for the grid and helps to ensure reliability in the case of higher than forecasted load (e.g., from extreme weather) or unforeseen power plant outages. This analysis assumes that regional reserve margins extant in 2024 are binding, i.e. that no system headroom exists, so that any increase in peak load requires new capacity investment. Thus, to supply additional load without building new capacity, there are only two options to avoid violating the minimum reserve margin constraint: (1) deploying demand response to avoid increasing the peak load; or (2) increasing the contribution of energy storage to reserve capacity by increasing discharge in peak hours. This assumption is central to the results: a region in which the 2024 reserve margin exceeded its requirement would have headroom that this framework assigns a value of zero. A complementary framework for assessing the extent of slack in the system is described in [15].

### *Representation of new large load and flexibility*

For the hourly profiles of new large loads, EPRI has collected load shapes for data centers with specific workloads (e.g., cloud services and enterprise) [16]. Although characteristics vary across workloads, data centers tend to operate with high load factors to minimize downtime of computational capacity. In the absence of generic load profiles that capture the average hourly shape of emerging large loads, we assume that new large loads have flat (i.e., constant over time) load profiles in this study. The term "large load" covers new loads above roughly 10 MW with high load factors, including but not limited to data centers.

Large load flexibility is represented as a demand response (DR) option with a very high dispatch cost of $100,000/MWh. This cost is intentionally set very high and does not represent the actual cost or value of DR but rather is used to ensure that the DR mechanism will only be deployed when existing generation and energy storage capacity are unable to meet the load in a given hour. The stylized DR option represents generic curtailment of grid-supplied power, e.g. through reducing compute or substituting grid-supplied power with on-site power. DR dispatch cannot exceed the new large load allocated to a given region and is assumed to be firm capacity that contributes to the reserve margin (i.e., DR allows new load to be added without requiring new capacity to meet the reserve margin). No cost, emissions, or availability constraint is explicitly attributed to on-site resources and is left for future work. The potential for other flexible demand resources (e.g., interruptible industrial loads, electric vehicle charging) to create headroom for new large loads is also left for future work.

While dispatchable technologies can be fully dispatched up to their net summer capacity in any given hour, upper bounds are imposed on the annual capacity factors of these technologies to reflect unforced outages and planned maintenance. The upper bounds for natural gas, coal, hydrogen, biomass, and geothermal technologies are set at 90%, while the annual capacity factors for nuclear technologies cannot exceed 95%. These technologies typically do not approach these upper bounds as actual utilization is constrained by economic dispatch (i.e., merit order). However, the upper bounds on the annual capacity factors may become binding as new large loads are introduced without allowing new capacity investments. Unit commitment, ramp rates, minimum stable output, startup costs, and fuel deliverability are not explicitly represented.

*Scenarios*

Two groups of scenarios are compared in the analysis. For each group, the model allocates a target national level of new large load (ranging from 10 to 150 GW) endogenously across regions, which provides insight into where deployment can best leverage system headroom and existing resources. In the first set of scenarios ("FixCap"), the generation and energy storage capacity is fixed at 2024 levels, so that all load must be supplied by existing assets and DR. In these scenarios, DR dispatch provides information on the characteristics and total amount of curtailment that is required to accommodate each increment of new large load.

A second set of scenarios ("OptCap") include the same levels of new large load identified in the FixCap scenarios, but investment in new generation and energy storage capacity is allowed, avoiding the DR option. The OptCap scenarios use the regional allocations of new large load identified in the associated FixCap scenarios so that differences between the OptCap and FixCap scenarios are directly comparable.

A separate pair of scenarios addresses announced siting of new data center loads. EPRI's medium 2030 data center projection assumes that 100% of under construction projects and 75% of planned projects in advanced stages materialize whereas only 10% of planned projects in early stages of development materialize, resulting in about 71 GW of realized peak load from data centers in 2030 of which 49 GW are new between 2024 to 2030 [1]. The Announced scenario fixes the regional distribution of that 49 GW to the projection while the Flexibility-Aligned scenario allocates the same national total to minimize required DR. Both are run under FixCap conditions.

All scenarios include current federal and state energy policies, including the One Big Beautiful Bill Act (OBBBA) [17].

*Curtailment metrics*

For each region and scenario, we report six metrics derived from the hourly DR dispatch: annual curtailed energy as a percentage of the annual energy of the new load; maximum curtailment depth as a percentage of regional new large load in the deepest hour; curtailment frequency as the number of hours with non-zero curtailment, also reported at depth thresholds of 25%, 50%, 75% and 90%; the number of independent curtailment events; and the mean and maximum duration of those events. National figures are load-weighted means across regions.

For assessing price impacts of relying on existing capacity, the analysis assumes that the costs of DR are allocated to large loads. Thus, from the perspective of the bulk power system, DR is simply seen as a reduction in load. Electricity prices are reported as the energy component of the wholesale price, in 2023 USD/MWh, taken as the load-weighted marginal energy cost. $CO_2$ emissions are power-sector combustion emissions and exclude emissions from any behind-the-meter generation used to deliver curtailment, and exclude emissions embodied in inter-regional imports, which are accounted to the exporting region.

*Caveats*

US-REGEN includes bulk transmission between model regions but does not represent transmission or distribution within regions. A recent EPRI study found that transmission bottlenecks significantly reduce system headroom [14]. This analysis also does not include plant-level operational constraints (e.g., unit commitment) that may shape dispatch in critical hours. The results from this study reflect available headroom based on existing generation and energy storage capacity as of 2024 and a stylized representation of regional reserve capacity requirements, excluding transmission and operational constraints that may further limit system headroom. Carbon-free electricity procurement requirements are not imposed, non-large-load demand growth is not represented, regional reserve margins are assumed to be exactly binding in 2024, and the new load is assumed perfectly flexible up to 100% depth. Each of these simplifications increases estimated headroom, so the values reported here should be interpreted as upper bounds.

**Acknowledgments**

The views and opinions expressed in this paper are those of the authors alone and do not necessarily reflect those of EPRI or its members. We thank Nidhi Santen for comments on an earlier draft.

**Supplementary Information**

*Supplementary Note 1: Peak Load and Reserve Margin Effects*

Figure S1 shows changes in regional absolute and net peak loads relative to 2024 at each level of new large load introduced nationally. The reserve margin constraint is assumed to be binding in all regions and wind, solar, and energy storage are assumed to contribute their actual generation/discharge to reserve capacity in a given hour. The reserve margin constraint is thus binding in the net peak hour in which load net of renewable generation is largest. It is possible for the absolute peak load to increase in regions in which energy storage discharge can be increased and/or renewable output aligns with the absolute peak. However, any increase in the absolute peak load increases the reserve requirement and thus the net peak load must decline in scenarios in which new capacity cannot be built. The decline in the net peak load can be achieved by discharging additional energy storage and/or curtailing load in the net peak hour. Thus, flexibility-enabled headroom is not simply a restatement of existing reserve surplus.

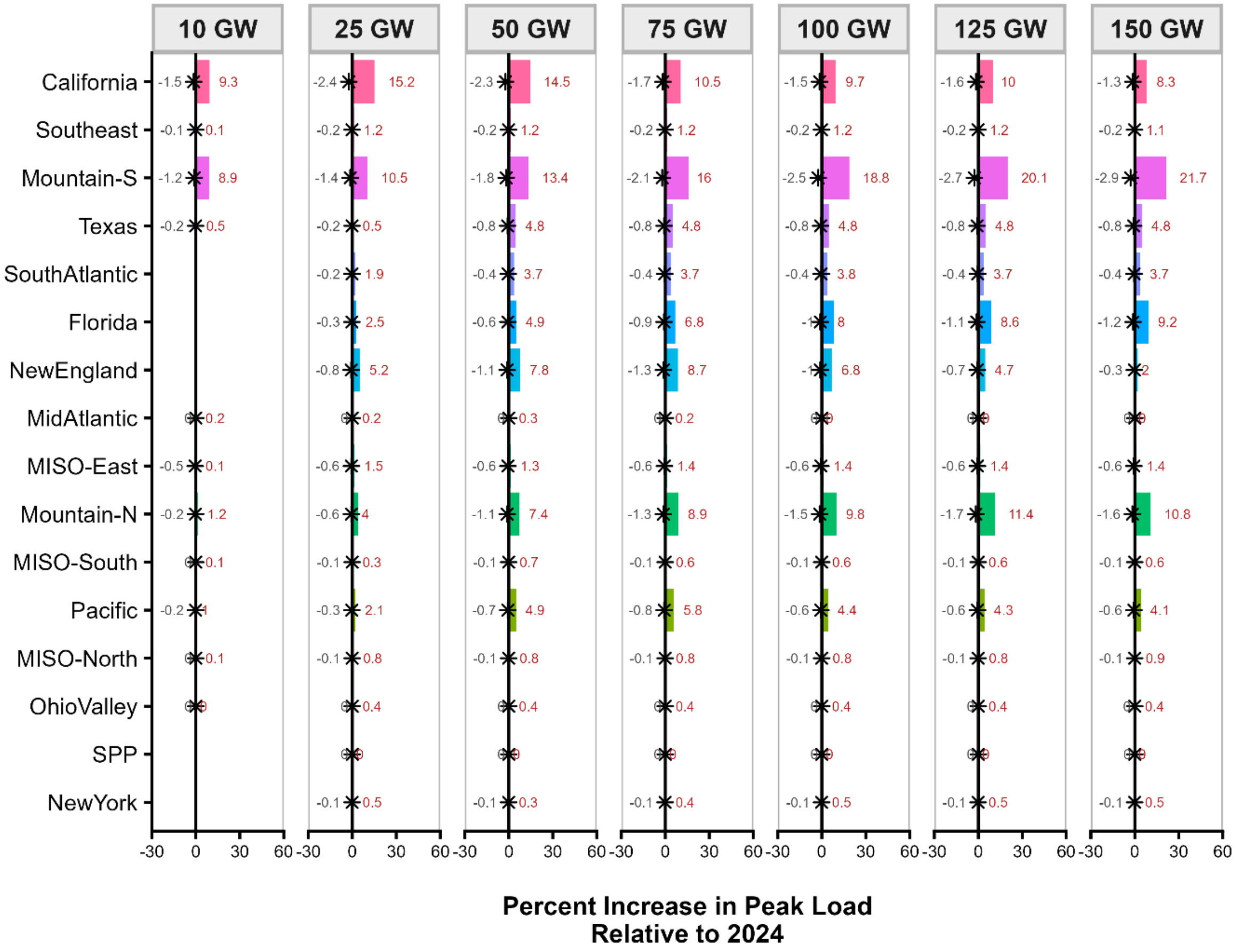

**Figure S1. Peak load and reserve margin effects.** (a) Percentage change in regional absolute and net peak loads relative to 2024 at each national new large load level. Positive values labeled in red are changes in absolute peak load and negative values indicated by a black asterisk and labeled in black are changes in net peak load.

*Supplementary Note 2: Costs*

Figure S2 illustrates incremental annualized system costs by component. Across the new load range examined, FixCap system cost is lower than OptCap, because the capital expenditure avoided by not building exceeds the additional fuel cost incurred by dispatching less efficient existing generators. The two curves cross only at higher load levels beyond the range shown.

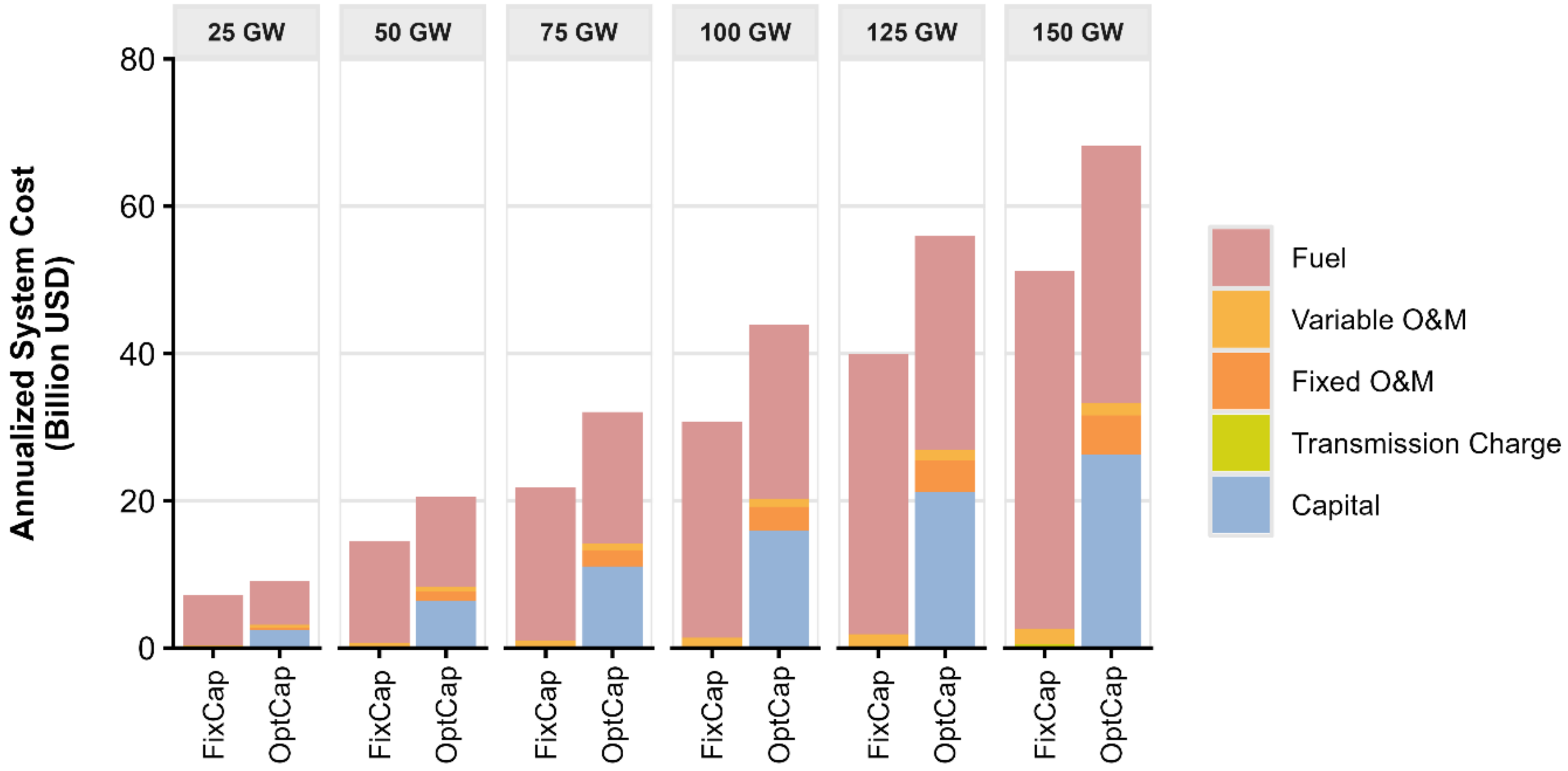

**Figure S2. Annualized incremental system costs across new large load levels.** Costs by component are shown under scenarios assuming fixed capacity (FixCap) and endogenous investment (OptCap) at each level. Costs incurred by the large load to deliver curtailment (e.g., behind-the-meter capacity) are not included in the analysis.

Several caveats are important for contextualizing these results, which primarily concern cost categories that are absent from the FixCap bars. First, costs associated with providing curtailment (which Figure 3 shows must equal its full peak load in many regions) are not included. These costs could include lost revenue from reduced compute and/or the costs of building and operating any on-site energy resources that could be used to provide DR while avoiding reduction in compute. Second, emissions are not explicitly priced. Third, any cost of the reliability risk transferred to the load is not represented. Finally, these costs are for a single period only and do not reflect longer-run changes, including lower costs over time for the OptCap scenario once capital-intensive, low-marginal-cost resources have been added.

*Supplementary Note 3: Power System Impacts in Regions with Large Announced Data Center Loads*

Figure S3 decomposes the regional consequences of serving announced data center load in the South Atlantic region and Texas, which account for 40% of projected new data center loads between 2024 and 2030.

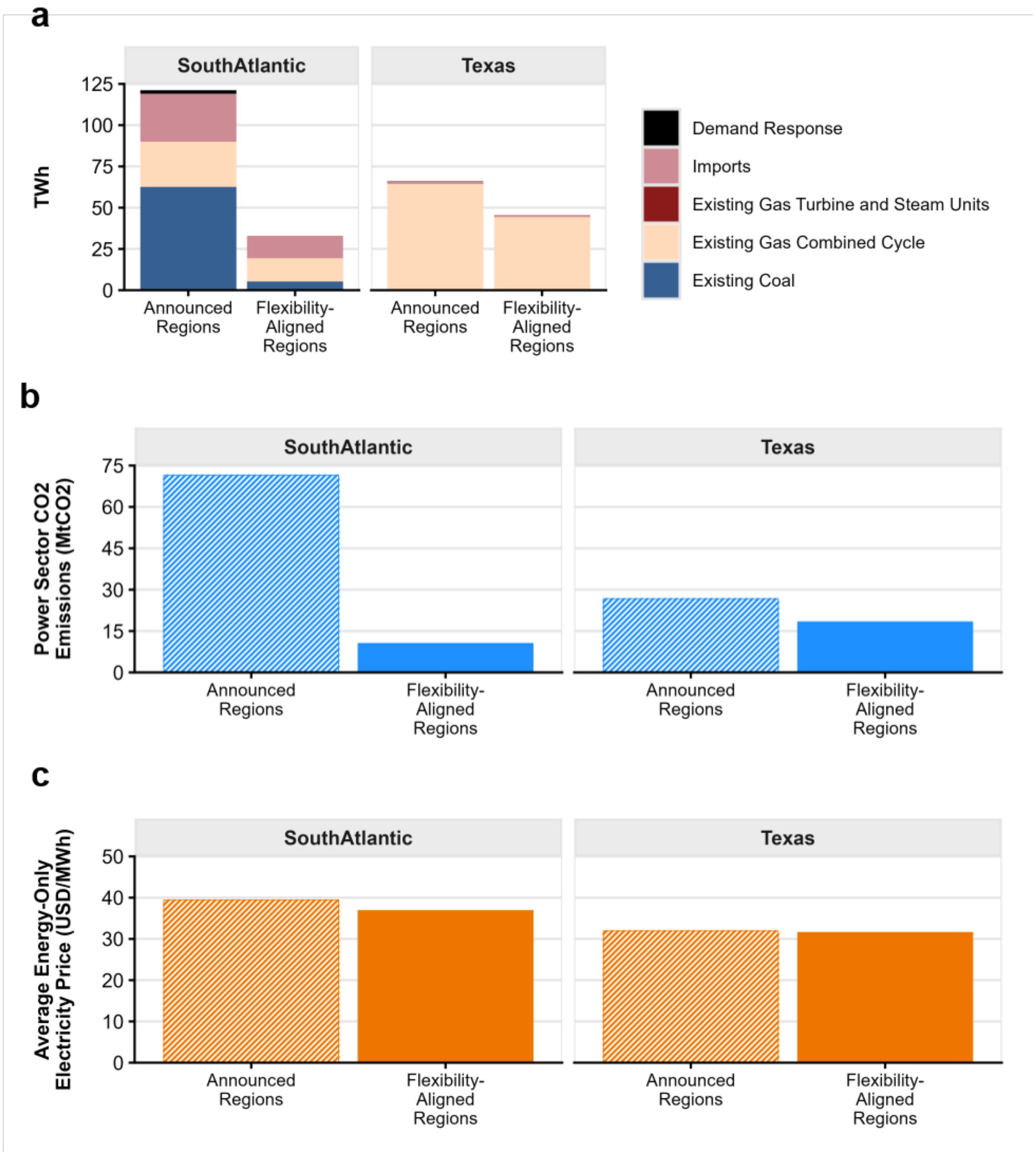

**Figure S3. Regional power system impacts of announced versus flexibility-aligned siting in South Atlantic and Texas.** (a) Incremental generation and imports. (b) Incremental regional power-sector $CO_2$ emissions. (c) Regional average energy-only wholesale electricity price. Incremental quantities are relative to a scenario with no new data center load.

South Atlantic cannot supply the announced load from within the region. About 75% of the locally served energy comes from increased capacity factors of existing coal and gas units, and the balance requires roughly 29 TWh of additional imports relative to a no-new-load baseline. Because imported energy carries emissions that are accounted to the exporting region, the reported regional emissions understate the true increase. This leakage effect is one that any region-level carbon accounting applied to data center load will systematically miss, and the leakage effect becomes more pronounced as a constrained region leans increasingly on imports. In South Atlantic, regional electricity $CO_2$ emissions are roughly 70 $MtCO_2$ under announced siting against roughly 10 $MtCO_2$ under flexibility-aligned siting, and the energy component of the regional wholesale price is about 7% higher.

In Texas, existing gas combined-cycle units serve the incremental load under both allocations, so the marginal generator (and therefore the price) is largely unchanged, with the energy component about 1% higher under announced siting. About 0.2% of the annual energy of the new load must still be curtailed to satisfy the reserve margin.

*Supplementary Note 4: Carbon-Free Electricity Procurement*

All scenarios in the analysis allow the incremental load to be served by the economic mix of generation, energy storage, and transmission, which in FixCap cases means predominantly existing coal and gas. Operators with carbon-free electricity commitments would face a materially different problem: the regions with the most flexibility-enabled headroom are not in general those with surplus carbon-free generation, so a CFE requirement would tighten siting further and reduce headroom below the values reported here. A qualification applies in the opposite direction. Load that procures renewable output may displace that output from other consumers rather than change total fossil generation, so the system-level emissions effect of a procurement requirement is not the same as its facility-level effect [18]–[20].

*Supplementary Note 5: Extended Caveats*

The following limitations are stated in Methods and are expanded here:

- Intra-regional networks: US-REGEN represents bulk transfers between its 16 regions but does not include explicit constraints on transmission or distribution within them, though costs for intra-regional lines are included. A separate assessment finds transmission bottlenecks reduce headroom substantially [14], which is likely the largest upward bias in our estimates.
- Plant-level operations: Unit commitment, minimum stable output, ramp rates, start costs, forced outage events and fuel deliverability are not represented. Each tightens the critical hours in which curtailment depth is determined.
- Weather and load years: A single weather year (2024) supplies load, renewable output, and the calibration of energy storage capacity value. Correlated extremes (e.g., a wide-area heat event with low wind) are conditions under which depth requirements bind and are not sampled.
- Load profiles: New load is assumed to be flat. AI training and inference loads are neither flat nor identical to one another, and generic hourly shapes are not yet available [16].
- Other flexible demand: Industrial process load, managed electric vehicle charging, and flexible electrified end uses could supply headroom for large loads but are not represented.
- Non-large-load growth: Demand growth from electrification outside the large-load category is excluded, so the available headroom is not shared with other new demand.
- Curtailment cost and value: Curtailment is priced administratively, not economically. No supply curve for flexibility is represented, and no cost is attributed to on-site resources.

These simplifications bias estimates in the same direction and imply that the headroom values reported in this paper are upper bounds.

**Table S1. Scenario definitions**

| Scenario family | New large load (GW) | Capacity treatment | Regional allocation |
|---|---|---|---|
| FixCap | 10, 25, 50, 75, 100, 125, 150 | Fixed at 2024 levels | Optimized to minimize curtailment |
| OptCap | 10, 25, 50, 75, 100, 125, 150 | New investment permitted | Adopted from matching FixCap case |
| Announced | 49 (2030 projection) | Fixed at 2024 levels | Fixed to announced project locations |
| Flexibility-Aligned | 49 (2030 projection) | Fixed at 2024 levels | Optimized to minimize curtailment |